# A Novel Hierarchical Multi-Agent System for Payments Using LLMs

Joon Kiat CHUA[1,†][0009−0001−8333−9792], Donghao HUANG[1,2,†][0009−0005−6767−4872], and Zhaoxia WANG[1,*][0000−0001−7674−5488]

[1] School of Computing and Information Systems, Singapore Management University, 80 Stamford Rd, Singapore 178902, Singapore
[2] Research and Development, Mastercard, 4250 Fairfax Dr, Arlington, VA 22203, USA

**Abstract.** Large language model (LLM) agents, such as OpenAI's Operator and Claude's Computer Use, can automate workflows but unable to handle payment tasks. Existing agentic solutions have gained significant attention; however, even the latest approaches face challenges in implementing end-to-end agentic payment workflows. To address this gap, this research proposes the Hierarchical Multi-Agent System for Payments (HMASP), which provides an end-to-end agentic method for completing payment workflows. The proposed HMASP leverages either open-weight or proprietary LLMs and employs a modular architecture consisting of the Conversational Payment Agent (CPA - first agent level), Supervisor agents (second agent level), Routing agents (third agent level), and the Process summary agent (fourth agent level). The CPA serves as the central entry point, handling all external requests and coordinating subsequent tasks across hierarchical levels. HMASP incorporates architectural patterns that enable modular task execution across agents and levels for payment operations, including shared state variables, decoupled message states, and structured handoff protocols that facilitate coordination across agents and workflows. Experimental results demonstrate the feasibility of the proposed HMASP. To our knowledge, HMASP is the first LLM-based multi-agent system to implement end-to-end agentic payment workflows. This work lays a foundation for extending agentic capabilities into the payment domain.

## 1 Introduction

The emergence of autonomous LLM-based agents such as OpenAI's Operator [14] and Anthropic's Claude Computer Use [2] marks a transformative moment in automating workflows with natural language. This has also attracted

---

attention in using agents for Agentic Payments and Commerce, where the goal is to have LLM agents enhance our online shopping experiences [11]. In research, there have been several studies in the use of both single agent and LLM-based multi-agent systems (MAS) for commerce use-cases, but they do not implement end-to-end agentic payments [5, 7, 17].

The root causes of this are multifaceted. Traditional payment network infrastructure employs anti-bot mechanisms, multi-factor authentication (e.g., 3D-Secure), and fraud detection systems as controls that places adoption constraints on agentic technology [11, 4]. Additionally, regulatory frameworks such as PCI DSS impose stringent requirements on any system that handles payment data, with guidelines now explicitly governing the use of AI in payment environments [15]. This complicates integrations, where merchant agents must facilitate payments without exposing sensitive payment data. Moreover, inherent hallucination risks in LLMs pose a challenge in LLM-based agentic use-cases [8, 10, 24]. In response, industry has explored agentic payment schemes that centers around tokenized digital credentials and onboarding of existing agentic systems onto traditional payment platforms [13, 20]. While these approaches represent a step forward, they require non-trivial changes to traditional payment rails, which introduces integration and structural overheads between payments and agentic workflows [11].

These challenges also reveal a fundamental limitation in current research. Despite the advances in LLMs and LLM-based multi-agent capabilities, there is still no established reference framework for achieving end-to-end agentic payments using LLM-based agents [8, 16, 19]. Therefore, to bridge this gap in end-to-end agentic payments, we aim to address the question: Can we design a framework that enables external LLM agents to converse and complete payments using natural language with minimal integration overhead?

In this paper, we propose a novel Hierarchical Multi-Agent System for Payments (HMASP). HMASP modularizes payment processing within a hierarchical multi-agent setup. At its core is the Conversational Payment Agent (CPA), a conversational intermediary that serves as a central entry point, which external agents can interface with using natural language to invoke payment-related tasks. HMASP's incorporate architectural patterns that enable modular task-execution for payment-related processing across different agents. Experiment results demonstrate the feasibility of HMASP in enabling end-to-end agentic payments using both open-weight and proprietary LLMs. The main contributions of this work are as follows:

- **Novel Hierarchical Multi-Agent Payment Framework:** We propose a novel Hierarchical Multi-Agent System for Payments (HMASP) using LLMs, laying the foundation for LLM-based multi-agent systems to achieve end-to-end agentic payment workflows.
- **Modular Design for Payment Execution:** The proposed HMASP adopts a modular design comprising the Conversational Payment Agent (CPA), Supervisor agent, Routing agent, and Process summary agent, with the CPA serving as the core component of the architecture by handling all external requests and coordinating subsequent tasks across hierarchical levels, thereby enabling modular task execution distributed across multiple agents.
- **Compatibility with LLMs:** The proposed HMASP can leverage either open-weight or proprietary LLMs, making it flexible and adaptable across different AI platforms.
- **Demonstrated Feasibility**: Experimental results show that HMASP can successfully implement end-to-end agentic payment workflows, within a simulated environment, marking it as the first LLM-based multi-agent system in this domain.

## 2    Related Works

**LLM-based Multi-Agent Systems.** Recent research has highlighted the various structures and strategies that enable LLM-based multi-agent cooperation and collaboration across domains [8, 19]. A notable strategy is the use of role-based protocols, organizing agents into specializations [8, 9, 19, 21], which has shown better collaborative performance as demonstrated in coding tasks such as MetaGPT [9]. Nonetheless, the risk of LLM hallucinations pose a challenge to agentic reliability [8, 10, 24]. Addressing this, Cemri et al., (2025) demonstrated in their paper on Multi-Agent System Failure Taxonomy that LLM-based MAS failures often stem from poor system design, and not just model reliability. This emphasizes that architectural design is key to ensuring reliability in agentic workflows [3]. Building on these insights, HMASP leverages on role-based agents and introduces several architectural patterns to enable task modularization, to achieve end-to-end agentic payments.

**LLMs for Payments.** There have been several studies on using LLM agents in payment-related domains [5, 7, 17]. These include using LLM as a virtual assistant to help users navigate mobile payment applications like AliPay [7] and using LLMs to enhance fraud and scams detection for payment transactions [5, 17]. While this demonstrates LLM's agentic capabilities, current use-cases stop at the payment network boundary. This is because traditional payment rails impose technical barriers such as fraud and authorization controls that limit adoption of agentic technology [4, 11]. As such, there is a lack of reference framework on how to achieve end-to-end agentic payments  [8, 16, 19]. To overcome this gap, we propose a hierarchical LLM-based multi agent system (HMASP) that modularizes payment processing to enable end-to-end agentic payments.

## 3    Methodology

This section details the methodology on the proposed Hierarchical Multi-Agent System for Payments (HMASP) to complete end-to-end agentic payment-related tasks. HMASP uses LangGraph [12] as the orchestration framework.

### 3.1   Layout of the proposed HMASP

In our HMASP architecture, agents are organized into roles based on function and specialization [8, 9, 19, 21]. Agents were prompted on their roles, expected responses, possible handoffs, and tasks that are within their scope. As shown in Fig. 1 four primary agent types are defined in our framework; the Conversational Payment Agent, Supervisor agent, Routing agent, and Process summary agent.

**Conversational Payment Agent (CPA) (First Agent Level).** The CPA serves as the entry point into our HMASP and handles all external requests. This agent coordinates tasks between all Supervisors by ensuring proper handoffs based on the input's relevance. Additionally, it is also responsible for generating replies to the request. CPA can also reject irrelevant requests.

**Supervisor Agent (Second Agent Level).** The Supervisor agent serves as the decision-maker and orchestrator of all domain-specific workflows (e.g., payments-related domain, cards-related domain). Each Supervisor is in charge of delegating requests to a task-specific workflow (e.g., Card Registration) within each domain for further handling. The Supervisor is also responsible for reporting back to the CPA on the outcomes of their task-specific workflow, generated by Process summary agents, for further handling. Supervisors can also reject irrelevant requests.

**Routing Agent (R) (Third Agent Level).** Within each task-specific workflow, the Routing agent receives input from Supervisors. They serve as the final decision layer on whether to trigger the task-specific workflow, or reject irrelevant requests and route to a Process summary agent for handling.

**Process Summary Agent (P) (Fourth Agent Level).** Process summary agents generate responses on the outcomes of their associated task-specific workflows to send back to Supervisor agents. These are structured responses containing the key information of each process workflow, and a summary of the success or failure of the workflow. They can also generate rejection responses on behalf of Routing agents.

**Structured Handoffs between Agents.** To enable cooperation between the agents, we implemented structured handoffs using LangGraph. Handoffs function as a way for agents to seamlessly pass processes to one another, where an agent can invoke the downstream or upstream agents [12]. We use this to enable coordination where agents handoff the external request sent sent to the CPA till it reaches the right agent or workflow for handling. Agents will decide where to handoff to based on the input. If handoffs are not required, such as when receiving an irrelevant input, agents will generate a response to reject the request.

Fig. 1 illustrates the overall architecture and workflow of the proposed method. The process begins when an external request is received by the Conversational

Payment Agent (CPA - first agent level) (Fig. 1(a)), which acts as the primary orchestrator. The CPA delegates the request to one of two available Supervisor agents (Fig. 1(b1) or (b2) - second agent level) based on the request's context. The selected Supervisor then forwards the request to an appropriate Routing agent (Fig. 1(c1), (c2), or (c3) - third agent level), which identifies and initiates the corresponding task-specific workflow (Task 1, Task 2, or Task 3). Once initiated, a series of function modules within each task-specific workflow are triggered to complete the request. Once the task is completed, the Process summary agent (Fig. 1(d1), (d2), or (d3) - fourth agent level) generates an outcome summary, which is then propagated back through the system. If user input (Fig. 1(e)) is needed during workflow execution, an interrupt mechanism is triggered to pause the workflow and collect the necessary input before proceeding. Finally, the CPA delivers a response (Fig. 1(f)), confirming successful request completion.

### 3.2   States and Information Processing

**States for Information Processing.** A Graph State [12] consists of user-defined state variables that persist and can be retrieved throughout the MAS runtime. We defined three different role-based states that are used by the different agents and payment processing modules - (1) `PayAgentState`: Used solely by the CPA, (2) `CardsState`: Used by all agents and modules involved in card-related tasks and (3) `PaymentState`: Used by all agents and modules involved in payments-related tasks. Agents and modules can only access their assigned states, preventing unauthorized read or write across state variables. To enable information sharing, however, we define several shared state variables (e.g., user ID, card ID) that span multiple states. Only these shared variables are visible across states. This ensures modularity, where only necessary information is propagated and all other sensitive information is isolated.

**Decoupled Message States.** Similar to agent states, our framework decouples the message state to be role-based. Agents can only invoke their own message states to generate responses. An agent's message state is updated in three ways - **(i) upstream messages:** these are incoming messages containing the response from upstream agents, **(ii) self-outputs:** generated responses derived from the agent invoking its own message state, **(iii) downstream-responses:** summary responses propagated back from downstream agents or workflows. By decoupling message states, agents can only interpret role-specific messages, minimizing token usage and reduces exposure to unnecessary or sensitive information from other agents or workflows in the framework. We also appended the agent's name (e.g. "CPA: message") to their responses for role-based attribution.

**Payment Related Workflows.** In Fig. 1, within each payment related workflow, Routing agents and Process summary agents are connected to function modules essential for task completion (e.g., Authorize 3DS in Fig.1 Task 3). These function modules are connected as a LangGraph subgraph [12], which can be invoked by the Routing agents. At the end of the execution, control is passed to the processing agent for further handling. Thus, payment related function modalities are isolated within their own subgraph enforcing security by design, as all sensitive operations are isolated within their workflows.

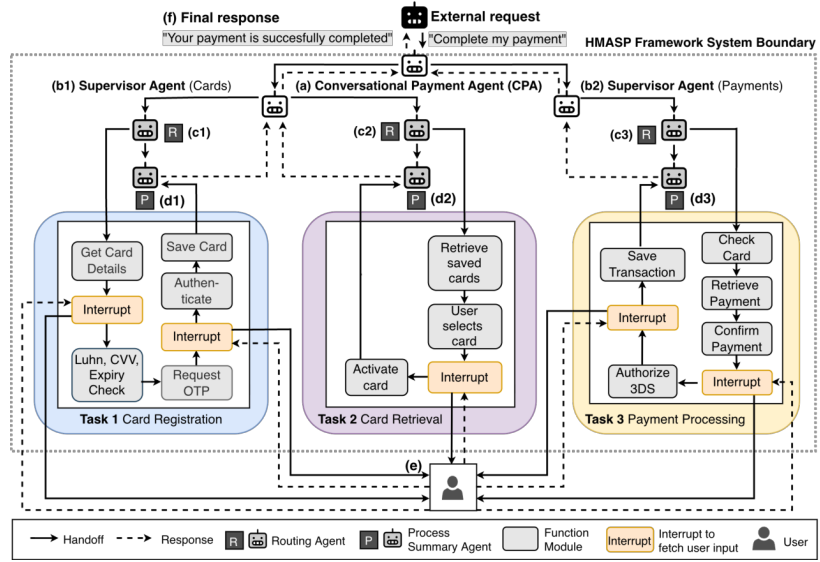

**Fig. 1.** Overview of the proposed HMASP. An external request (e.g., "Complete my payment") is sent to the Conversational Payment Agent (CPA - first agent level); (a) The CPA processes the request and delegates it to one of the two Supervisors (b1, b2 - second agent level) for further handling; (b) The selected Supervisor routes the request to a suitable Routing agent (c1, c2, or c3 - third agent level); (c) Routing agents initiate the corresponding task-specific workflow (Task 1, Task 2, Task 3). Upon initiation, a series of task-relevant function modules are triggered to complete the request; (d) Upon task completion, the Process summary agent (d1, d2, d3 - fourth agent level) compiles the outcome and generates a response. This summary response is propagated upward the system hierarchy through Supervisor agents and CPA; (e) When user input is required to complete the task, an interrupt mechanism is triggered to obtain the necessary input before execution resumes; (f) Finally, the CPA responds to the request (e.g., "Your payment has been successfully completed).

### 3.3   Reliable Execution Through Determinism

We introduced design patterns to decouple key payment workflows from LLM-generated content to ensure determinism and mitigate against hallucination risks when handling key payment information using LLM-based agents.

**Interrupt (Human-in-the-Loop).** A critical component of payment processing would be gathering sensitive information from users (e.g., card details). We utilized LangGraph's `interrupt` mechanism for persistent and reliable human-in-the-loop within our framework [12]. As shown in Fig.1(e), to fetch user input, `interrupt` is triggered pausing the workflow execution and raising a request. Users can then respond in natural language, and the respond goes through a series of validation and checks (e.g. 16 digit card number checksum - Luhn Check). Upon successful validation, the workflow resumes and the user input information is stored in state variables. If any checks fail, the workflow stops and is handed off to the Process summary agent for exception handling.

**State Variables for Determinism.** User input information and certain workflow outputs are stored in state variables to enforce information determinism and persistence. When this information is needed, rather than relying on LLM-agents to generate and pass the information, function modules and agents will retrieve this from the state variable. Therefore, critical information needed to complete workflows can never be tampered or hallucinated. This decouples payment-related processing from generative processes ensuring information determinism.

### 3.4   Evaluation Dataset

To the best of our knowledge, while many datasets exist for other LLM-based multi-agent simulations, none are specifically designed for conversational payment processes. This work is led by Mastercard, a Global Technology Company in the Payments Industry, with deep expertise in payment technology. The dataset used is part of Mastercard's research, and can be made available based on request. It includes a total of 1000 data points, distributed across four different tasks (categories) with 250 data points per task:

1. **Task 1 - Relevant Inputs for Card Registration**: Inputs relevant to registering a new card, which will trigger the card registration workflow (e.g., "Register a new card")
2. **Task 2 - Relevant Inputs for Card Retrieval**: Inputs relevant to checking saved cards, which will trigger the card retrieval workflow (e.g.,"List my cards").
3. **Task 3 - Relevant Inputs for Payment Processing** : Inputs relevant to checkout/payment, triggering the payment processing workflow (e.g.,"Take me to payment checkout").
4. **Task 4 - Irrelevant Inputs**: Inputs that are obviously unrelated to payments and should not trigger any payment-related workflow (e.g.,"Can you tell me a joke?").

### 3.5   Evaluation Setup and Metrics

In this section, we detail the evaluation setup and metrics used to evaluate our HMASP performance across both open-weight and proprietary models.

**Evaluation Setup.** Payment modules were simulated using functions, reflecting the architectural decoupling of the agent layer from underlying payment APIs.

The evaluation assesses the end-to-end agentic workflow within a simulated environment, focusing on task success rate and agent handoff reliability. Production complexities (e.g., issuer authentication and PCI-DSS validation), are expected to affect latency rather than the architectural properties evaluated. We evaluated a set of open-weight models against GPT-4.1 [1] (2025-04-14 release) as the proprietary baseline. Open-weight models include Qwen models [22, 23] (Qwen3:8b, Qwen3:14b, Qwen3:32b, Qwen2.5:7b, Qwen2.5:14b, Qwen2.5:32b), Mistral models [18] (Mistral-small3.2:24b, Magistral:24b), and Meta's Llama models [6] (Llama3.1:8b, Llama3.1:70b). All HMASP agents utillized the same model. Open-weight models were ran locally using Ollama[3] v0.9.6, on an Ubuntu 24.04.1 LTS system equipped with an NVIDIA H100 GPU with 96 GB VRAM.

**Task Success Rate.** Task success rate is defined as the proportion of runs that (i) trigger the correct workflow and (ii) correctly saving or retrieving the required information. For Tasks 1 to 3, we validated correct registration of card details, retrieving correct last 4 digits of selected card, and saving the correct transaction respectively. For Task 4 (irrelevant inputs), no workflow should be triggered.

**Agent Handoff (F1-Score).** Agent handoff was evaluated by comparing actual versus expected handoffs: (i) from CPA to the Supervisor, and (ii) from Supervisor to the relevant workflow processes. For relevant inputs (Tasks 1 to 3), we captured both (i) CPA-to-Supervisor and (ii) Supervisor-to-workflow handoffs. A true positive is defined as an actual handoff that matches an expected handoff. For irrelevant inputs (Task 4), no handoffs are expected. Thus, we captured only (i) CPA-to-Supervisor interactions. A true positive is defined as no handoff occurring, where the CPA directly rejects the input without triggering the Supervisor agent or any workflow.

## 4   Results and Discussion

**Task Success Rate.** Table 1 shows the overall task success rates across the four input tasks. GPT-4.1 delivers the best performance in our proposed HMASP method. Among open-weight models, Qwen2.5:32b shows relatively better task success rates ($\geq$95.6%) across all tasks. On the other hand, Qwen3, Llama and Mistral models are more inconsistent. For instance, while Qwen3:32b shows a 99.6% success rate on Task 1, it underperforms on Tasks 2 and 3 (35.2% and 39.6% respectively). Interestingly, from Table 1, Magistral:24b and Llama3.1-8b do not handle Tasks 1 to 3 well with $\leq$ 50% success rate, as the workflow stops either at the CPA or Supervisor level without triggering the correct task-relevant workflow. Similarly, Llama3.1-70b showed a low success rate on Task 3 for the same reasons. Lastly, the results demonstrated that most of the models can reject inputs that are obviously unrelated to payments (Task 4).

---

[3] `https://ollama.com/`

| Model | Task Success Rate (%) | | | |
|---|---|---|---|---|
| | **T1** | **T2** | **T3** | **T4** |
| Qwen3:8b | 84.0 | 90.4 | 52.4 | 99.6 |
| Qwen3:14b | 98.8 | 33.2 | 64.0 | 99.6 |
| Qwen3:32b | **99.6** | 35.2 | 39.6 | **100** |
| Qwen2.5:7b | 93.6 | 91.6 | 68.4 | **100** |
| Qwen2.5:14b | 94.0 | 90.4 | **96.4** | **100** |
| Qwen2.5:32b | 96.4 | **98.8** | 95.6 | **100** |
| Mistral-small3.2:24b | 84.0 | 90.8 | 47.2 | **100** |
| Magistral:24b | 44.0 | 18.4 | 3.2 | **100** |
| Llama3.1:8b | 6.0 | 10.4 | 4.0 | **100** |
| Llama3.1:70b | 66.0 | 74.0 | 26.0 | 98.8 |
| GPT-4.1 | **99.6** | **99.6** | **100** | **100** |

**Table 1.** Task success rate of open-weight and proprietary models in our proposed HMASP across four different tasks: payment-relevant tasks (T1, T2, and T3) and payment-irrelevant tasks (T4). Currently, all agents in our HMASP framework used the same LLM model, and this table illustrates the performance. Task Success Rate measures correct workflow triggering with task-relevant information saved.

**Agent Handoff Performance (F1-Score).** Results in Table 2 demonstrates the handoff performance across various agents. For relevant inputs (Tasks 1-3), GPT-4.1 achieved a high average F1-score of 99.9%, while open-weight models like Qwen2.5:32B achieving a comparable average F1-score of 98.9%. Most models correctly rejected irrelevant inputs (Task 4) at the CPA level without handing off to the Supervisor. Llama 3.1–8B and 70B, however, demonstrated weaker performance in agent handoff for Task 4. Unlike other open-weight models, the Llama3.1 models did not consistently reject irrelevant inputs at the CPA level and instead it still hands off the request to the Supervisor. These results highlight the limitations of Llama 3.1 models within our proposed method.

**Novelty Analysis.** Table 3 shows a comparative overview of existing existing autonomous agents in payment-relevant processes. To the best of our knowledge, no existing LLM-based multi-agent system support fully end-to-end agentic payment workflows. As a whole, results demonstrate the viability of leveraging several open-weight and proprietary models with our proposed HMASP method to achieve end-to-end agentic payment-related task processing.

## 5 Conclusion

In this work, we proposed a novel Hierarchical LLM-based Multi-Agent System for Payments (HMASP). Architecturally, HMASP comprises four hierarchical agent levels - the CPA, Supervisors, Routing, and Process summary agent. The system employs a modular architecture that leverages the CPA as a conversational intermediary to handle external requests and coordinate tasks. By utilizing shared state variables, decoupled message states, and structured handoff protocols, HMASP enables modular execution for payment-related tasks distributed across agents to achieve end-to-end agentic workflows. Experimental results demonstrate the feasibility of the proposed HMASP. This research shows that with HMASP, certain open-weight models (e.g., Qwen2.5:32b) can achieve performance comparable to GPT-4.1. To the best of our knowledge, there is no established reference framework for implementing end-to-end agentic payment workflows using LLM-based multi-agent systems. This underscores the novelty of HMASP in addressing this gap, laying the foundation for extending agentic capabilities into the payment domain.

| | Agent Handoff F1-Score(%) | | | |
|---|---|---|---|---|
| **Model** | **T1** | **T2** | **T3** | **T4** |
| Qwen3:8b | 93.2 | 79.4 | 73.0 | 99.8 |
| Qwen3:14b | 99.2 | 46.3 | 84.6 | **100** |
| Qwen3:32b | **99.9** | 55.7 | 66.6 | **100** |
| Qwen2.5:7b | 96.6 | 93.3 | 81.2 | **100** |
| Qwen2.5:14b | 95.0 | 92.0 | 98.9 | **100** |
| Qwen2.5:32b | 98.1 | **99.7** | **99.0** | **100** |
| Mistral-small3.2:24b | 94.0 | 93.7 | 84.8 | **100** |
| Magistral:24b | 64.9 | 67.9 | 63.7 | 99.8 |
| Llama3.1:8b | 65.5 | 65.7 | 68.4 | 11.4 |
| Llama3.1:70b | 90.2 | 91.1 | 77.3 | 44.4 |
| GPT-4.1 | **99.9** | **99.8** | 100 | 100 |

**Table 2.** Agent handoff performance evaluation using F1-Score for payment-relevant inputs (T1, T2, and T3) and payment-irrelevant inputs (T4). Agent handoff was evaluated by comparing actual versus expected handoffs. For T1 to T3, expected handoffs (true positives) are from (i) CPA-to-Supervisor and (ii) Supervisor-to-workflow. For T4, no handoffs from CPA are expected (true positive). Llama3.1 models have a lower F1-score for T4 as the CPA handoffs to the Supervisor most of the time, rather than immediately rejecting the input request.

| Method | Open-Weight LLM | Full Payment Processing | TR Success Rate (%) | TR Precision-Recall (%) | TR F1-Score (%) |
|---|---|---|---|---|---|
| LLMPA [7] | Yes | No[1] | 76.5 | – | – |
| Method 1 [5] | No | No | 93.3 | 85.7–30.2 | 44.77 |
| Method 2 [17] | Yes | No | – | N/A[2]–73.7 | – |
| **HMASP (Ours)** | **Yes** | **Yes** | **97.7** | **99.4–98.8** | **99.2** |

**Table 3.** A comparative overview of existing autonomous agents in payment-relevant processes with their Task-Relevant (TR) metrics. For HMASP, averages from the best-performing open-weight LLM (Qwen2.5:32b) were reported. To the best of our knowledge, this is the first implementation of an LLM-based MAS supporting full end-to-end agentic payment processing as such, direct comparisons are limited.

**Limitations and Future Work.** The current dataset helps validate research viability before production, however, it is relatively small and may not fully capture the complexity of all payment workflows and diverse user behaviors. Therefore, testing the proposed HMASP using a larger dataset, when available, will be an interesting direction for future work. In addition, the pursuit of agentic payments is a multi-faceted problem involving multiple research domains. Thus, future work includes security research on developing payment-centric guardrails for agentic payments, and enabling secure tokenization of sensitive information within state variables. Lastly, a comparative study of various agentic architectures and agentic protocols for payments processing can be conducted.